\documentclass[letterpaper,twocolumn,10pt]{article}
\usepackage{usenix}

\pdfoutput=1

\usepackage{booktabs} 
\usepackage{epsfig,endnotes}
\usepackage{comment}
\usepackage{url}
\usepackage{listings}
\usepackage{textcomp}
\usepackage{amssymb}
\usepackage{amsmath}
\usepackage{bm}
\usepackage{paralist}
\usepackage{subcaption}
\usepackage{algorithm}
\usepackage{algpseudocode}
\usepackage{graphicx}
\usepackage{color,soul}
\usepackage{authblk}
\usepackage{float}
\definecolor{lbcolor}{rgb}{0.9,0.9,0.9}

\begin{document}
\title{FlashR: R-Programmed Parallel and Scalable Machine Learning using SSDs}

\author[1]{\rm Da Zheng}
\author[1]{\rm Disa Mhembere}
\author[3]{\rm Joshua T. Vogelstein}
\author[2]{\rm Carey E. Priebe}
\author[1]{\rm Randal Burns}
\affil[1]{Department of Computer Science, Johns Hopkins University}
\affil[2]{Department of Applied Mathematics and Statistics, Johns Hopkins University}
\affil[3]{Department of Biomedical Engineering, Johns Hopkins University}

\maketitle

\begin{abstract}
R is one of the most popular programming languages for statistics and machine
learning, but the R framework is relatively slow and unable to scale to large
datasets. The general approach for speeding up an implementation in R is to
implement the algorithms in C or FORTRAN and provide an R wrapper. FlashR takes
a different approach: it executes R code in parallel and scales the code beyond
memory capacity by utilizing solid-state drives (SSDs) automatically. It
provides a small number of generalized 
operations (GenOps) upon which we reimplement a large number of
matrix functions in the R \textit{base} package. As such, FlashR parallelizes
and scales existing R code with little/no modification. To reduce data movement
between CPU and SSDs, FlashR evaluates matrix operations lazily, fuses
operations at runtime, and uses cache-aware, two-level matrix partitioning.
We evaluate FlashR on a variety of machine learning and statistics algorithms 
on inputs of up to four billion data points.
FlashR out-of-core tracks closely the performance of FlashR in-memory.
The R code for machine learning algorithms executed in FlashR
outperforms the in-memory execution of H2O and Spark MLlib by a factor of
$2-10$ and outperforms Revolution R Open by more than an order of magnitude.
\end{abstract}

\section{Introduction}

The explosion of data volume and the increasing complexity of data analysis
generate a growing demand for scalable statistical analysis and machine
learning tools that are simple and efficient.
Simple tools need to be programmable, interactive, and extensible, 
allowing scientists to encode and deploy complex algorithms. 
Successful examples include R, SciPy, and Matlab.  Efficiency dictates that
tools should leverage modern computer architectures, including scalable
parallelism, high-speed networking, and fast I/O from memory and solid-state
storage. The current approach for utilizing the full
capacity of modern parallel systems often uses a low-level programming
language such as C and parallelizes computation with MPI or OpenMP.
This approach is time-consuming and error-prone, and requires machine learning
researchers to have expert knowledge in the parallel programming models.
 

While conventional wisdom addresses large-scale data analysis and machine
learning with clusters
\cite{mapreduce,spark,systemml,tensorflow,petuum,graphlab}, recent works
\cite{flashgraph,gridgraph,Matveev17,hotos} demonstrate a single-machine solution
can deal with large-scale data analysis efficiently in a multicore
machine. The advance of solid-state drives (SSDs) allows us to tackle data
analysis in a single machine efficiently at a larger scale with a cheaper price.
Previous SSD-based graph analysis frameworks \cite{flashgraph, gridgraph, graphene}
have demonstrated the comparable efficiency to state-of-the-art in-memory graph
analysis, while scaling to arbitrarily large datasets. This work extends
these findings to matrix operations using SSDs for machine learning and
data analysis.


To provide a simple programming environment for efficient and scalable machine
learning, we present FlashR, an interactive R-based programming framework that
executes R code in parallel and
out-of-core automatically. FlashR stores large
vectors and matrices on SSDs and overrides many R functions in the R
\textit{base} package to perform computation on these external-memory vectors
and matrices.
As such, FlashR executes existing R code with little/no modification.
FlashR focuses on optimizations in a single machine (with multiple CPUs and
many cores) and scales matrix operations beyond memory capacity by 
utilizing solid-state drives (SSDs).  
Our evaluation shows that we can solve billion row, Internet-scale 
problems on a single thick node, which can prevent the complexity,
expense, and power consumption of distributed systems when they are
not strictly necessary \cite{hotos}.



To utilize the full capacity of a large parallel machine, we overcome
many technical challenges to move data from SSDs to CPU efficiently for matrix
computations,
notably the large performance disparities between CPU and memory and between
memory and SSDs. The ``memory gap'' \cite{Wilkes01} continues to grow, with 
the difference between CPU and DRAM performance increasing exponentially. 
There are also performance differences between
local and remote memory in a non-uniform memory architecture (NUMA), which are prevalent
in modern multiprocessor machines. 
RAM outperforms SSDs by an order of magnitude for both latency and throughput.

FlashR evaluates expressions lazily and fuses operations aggressively
in a single parallel execution job to minimize data movement. FlashR
builds a directed acyclic graph (DAG) to represent a sequence of matrix
operations and grows a DAG as much as possible to increase the ratio of
computation to I/O. When evaluating the computation in a DAG, FlashR
performs two levels of matrix partitioning to reduce data movement in
the memory hierarchy. FlashR by default materializes
only the output matrices (leaf nodes) of a DAG and keeps materialized results in
memory in order to minimize data written to SSDs. FlashR streams
data from SSDs to maximize I/O throughput for most computation tasks.


We implement multiple machine learning algorithms, including principal component
analysis, logistic regression and k-means, in FlashR. On a large parallel machine
with 48 CPU cores and fast SSDs, the out-of-core execution of these R implementations
achieves performance comparable to their in-memory execution, while significantly
outperforming the same algorithms in H2O \cite{h2o} and Spark MLlib
\cite{spark}. FlashR effortlessly scales to datasets with billions
of data points and its out-of-core execution uses a negligible amount of memory
compared with the dataset size. In addition, FlashR executes the R functions
in the R MASS \cite{mass} package with little modification and outperforms
the execution of the same functions in Revolution R Open \cite{rro} by more
than an order of magnitude.

We believe that FlashR significantly lowers the requirements for writing
parallel and scalable implementations of machine learning algorithms; it also
offers new
design possibilities for data analysis clusters, replacing memory with larger
and cheaper SSDs and processing bigger problems on fewer nodes.
FlashR is released as an open-source project at http://flashx.io/.


\vspace{-10pt}
\section{Related Work}
Basic Linear Algebra Subprograms (BLAS) defines a small set of vector and
matrix operations. There exist a few highly-optimized BLAS implementations,
such as MKL \cite{mkl} and ATLAS \cite{atlas}. 
Distributed libraries \cite{trilinos, petsc, elemental}
build on BLAS and distribute computation with MPI.
BLAS provides a limited set of matrix operations and requires
users to manually parallelize the remaining matrix operations.

Recent works on out-of-core linear algebra \cite{Toledo99, Quintana-Orti12}
redesign algorithms to achieve efficient I/O access and reduce I/O
complexity. These works are orthogonal to our work and can be adopted.
Optimizing I/O
alone is insufficient. To achieve performance comparable to state-of-the-art
in-memory implementations, it is essential to move data efficiently both from
SSDs to memory and from memory to CPU caches.

MapReduce \cite{mapreduce} has been used for parallelizing machine learning
algorithms \cite{Chu06}. Even though MapReduce simplifies parallel programming,
it still requires low-level programming.
As such, frameworks are built on top of MapReduce to reduce programming complexity.
For example, SystemML \cite{systemml} develops an R-like script language for
machine learning. MapReduce is inefficient for matrix operations because
its I/O streaming primitives do not match matrix data access patterns.

The Spark \cite{spark} is a distributed, in-memory framework that provides more
primitives for efficient computation and provides a distributed machine
learning library (MLlib, \cite{mllib}).
Spark is the most efficient distributed engine.

Distributed machine learning frameworks have been developed to train machine
learning models on large datasets. For example, GraphLab \cite{graphlab}
formulates machine learning algorithms as graph computation; Petuum \cite{petuum}
is designed for machine learning algorithms with certain properties such as
error tolerance; TensorFlow \cite{tensorflow} trains machine learning models,
especially deep neural networks, with optimization algorithms such as
stochastic gradient descent.

Efforts to parallelize array programming include Revolution R \cite{rro} and
Matlab's parallel computing toolbox, which offer multicore parallelism and
explicit distributed parallelism using MPI and MapReduce. Other works focus
on implicit parallelization. Presto \cite{presto} extends R to sparse matrix
operations in distributed memory for graph
analysis. Ching et. al \cite{Ching12} parallelize APL code by
compiling it to C. Accelerator \cite{accelerator} compiles
data-parallel operations on the fly to execute programs on a GPU.

OptiML \cite{optiml} is a domain-specific language for developing machine
learning in a heterogeneous computation environment such as multi-core
processors and GPU. It designs a new programming language and relies on
a compiler to generate code for the heterogenous environment.

\section{Design}

\begin{figure}
\centering
\includegraphics[scale=0.3]{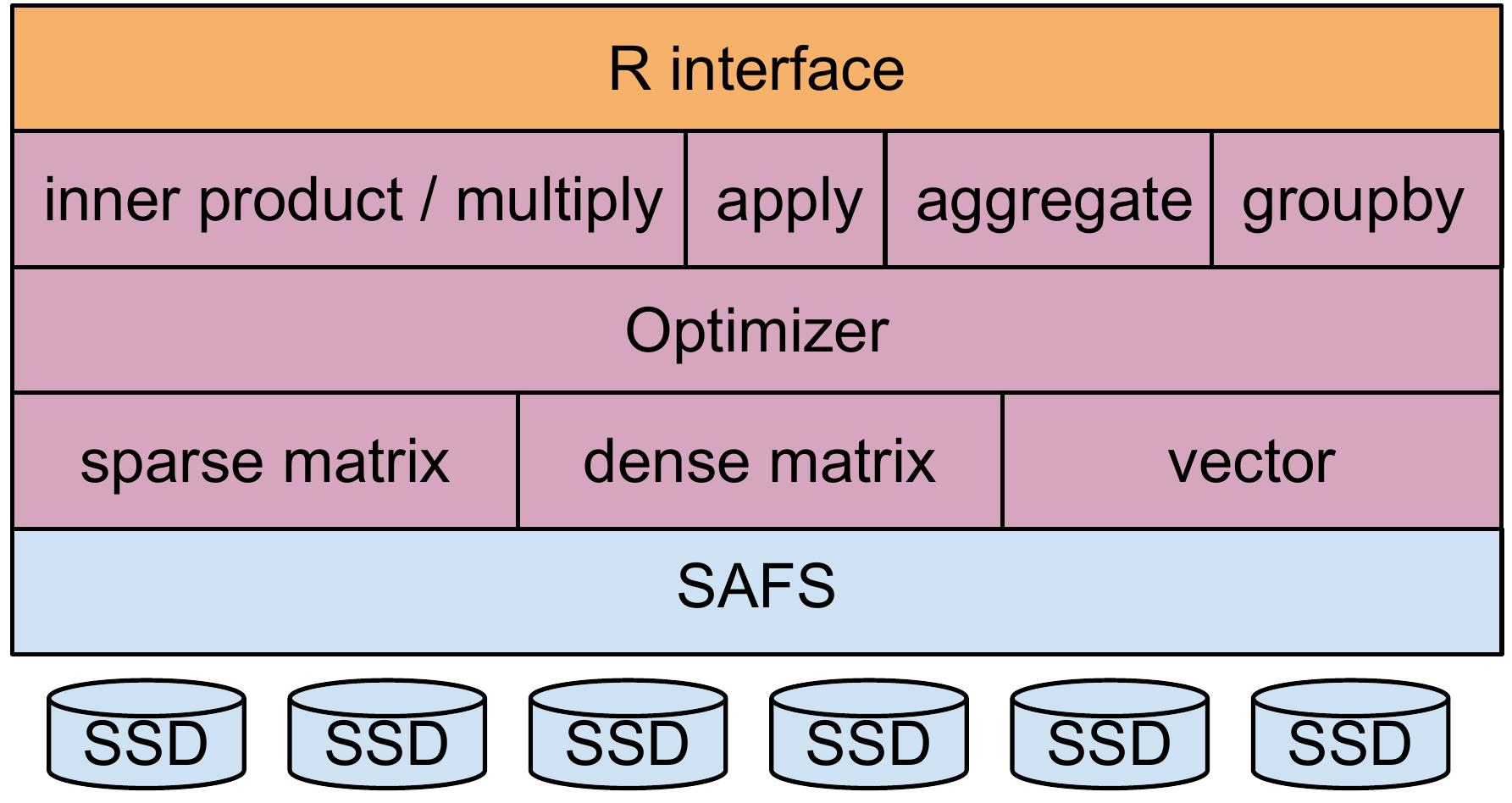}
\vspace{-5pt}
\caption{The architecture of FlashR.}
\label{fig:arch}
\vspace{-10pt}
\end{figure}

FlashR is a matrix-oriented programming framework for machine learning and
statistics. It scales matrix operations beyond memory capacity by utilizing
fast I/O devices, such as solid-state drives (SSDs), in a non-uniform memory
architecture (NUMA) machine. Figure \ref{fig:arch} shows the architecture of
FlashR. FlashR has only a few classes of generalized operations (GenOps) to
simplify the implementation and the GenOps improve expressiveness of the framework. 
The optimizer aggressively merges operations to
reduce data movement in the memory hierarchy.
It stores matrices on SSDs through SAFS \cite{safs},
a user-space filesystem for SSD arrays, to fully utilize high I/O
throughput of SSDs.

\subsection{Programming interface}

FlashR provides a matrix-oriented functional programming interface built
on Generalized Operations (GenOps).  GenOps (Table \ref{tbl:genops}) take matrices and
some functions as input and output new matrices that represent computation results.
The input function defines computation on individual elements in input matrices.
GenOps provide a flexible and concise programming interface and, thus,
we focus on optimizing the small set of matrix operations. All of
the GenOps are lazily evaluated to gain efficiency (Section
\ref{sec:datamove}).

\begin{table}
\begin{center}
\caption{Generalized operations (GenOps) in FlashR. $A$, $B$ and $C$ are
	matrices, and $c$ is a scalar. $f$ is a user-defined function that
	operates on elements of matrices.}
\vspace{-10pt}
\footnotesize
\begin{tabular}{|l|l|l|}
\hline
GenOp & Description \\
\hline
$C=sapply(A, f)$ & $C_{i,j}=f(A_{i,j})$ \\
\hline
$C=mapply(A, B, f)$ & $C_{i,j}=f(A_{i,j}, B_{i,j})$ \\
\hline
$C=mapply.row(A, B, f)$ & $C_{i,j}=f(A_{i,j}, B_j)$ \\
\hline
$C=mapply.col(A, B, f)$ & $C_{i,j}=f(A_{i,j}, B_i)$ \\
\hline
$c=agg(A, f)$ & $c=f(A_{i,j}, c)$, over all $i$, $j$ \\
\hline
$C=agg.row(A, f)$ & $C_i=f(A_{i,j}, C_i)$, over all $j$ \\
\hline
$C=agg.col(A, f)$ & $C_j=f(A_{i,j}, C_j)$, over all $i$ \\
\hline
$C=groupby(A, B, f)$ & $C_{k}=f(A_{i,j}, C_{k})$,\\ & where $B_{i, j}=k$, over all $i,j$ \\
\hline
$C=groupby.row(A, B, f)$ & $C_{k,j}=f(A_{i,j}, C_{k,j})$,\\ & where $B_i=k$, over all $i$ \\
\hline
$C=groupby.col(A, B, f)$ & $C_{i,k}=f(A_{i,j}, C_{i,k})$,\\ & where $B_j=k$, over all $j$ \\
\hline
$C=inner.prod(A, B, f1, f2)$ & $t=f1(A_{i,k}, B_{k,j})$,
\\ & $C_{i,j}=f2(t, C_{i,j})$, over all $k$ \\
\hline
\end{tabular}
\normalsize
\label{tbl:genops}
\vspace{-10pt}
\end{center}
\end{table}

GenOps are classified into four categories that describe different data access
patterns.

\noindent \textbf{Element-wise operations}:
\textit{sapply} is an element-wise unary operation; \textit{mapply}
is an element-wise binary operation; \textit{mapply.row} and
\textit{mapply.col} perform element-wise binary operations on
the input vector with every row or column of the input matrix
and output a matrix of the same shape as the input matrix.


\noindent \textbf{Aggregation}: \textit{agg} computes aggregation over
all elements in a matrix and outputs a scalar value; \textit{agg.row}
computes over all elements in every row and outputs a vector;
\textit{agg.col} computes over all elements in every column and
outputs a vector.

\noindent \textbf{Groupby}: \textit{groupby} splits the elements of a matrix
into groups, applies \textit{agg} to each group and outputs a vector;
\textit{groupby.row} splits rows into groups and applies \textit{agg.col}
to each group; \textit{groupby.col} splits columns into groups and applies
\textit{agg.row} to each group. Both \textit{agg.row} and \textit{agg.col}
output a matrix.

\noindent \textbf{Inner product} is a generalized matrix multiplication
that replaces multiplication and addition with two functions.


We reimplement a large number of matrix functions in the R \textit{base}
package with the GenOps to provide users a familiar programming interface.
By overriding existing R matrix functions, FlashR scales and parallelizes
existing R code with little/no modification. Table \ref{tbl:Rfuns} shows
a small subset of R matrix operations and their implementations
with GenOps.


\begin{table}
\begin{center}
\caption{Some of the R matrix functions implemented with GenOps.}
\vspace{-10pt}
\footnotesize
\begin{tabular}{|l|l|l|}
\hline
Function & Implementation with GenOps \\
\hline
$C=A+B$ & $C=mapply(A, B, "+")$ \\
$C=A-B$ & $C=mapply(A, B, "-")$ \\
$C=pmin(A,B)$ & $C=mapply(A, B, "pmin")$ \\
$C=pmax(A,B)$ & $C=mapply(A, B, "pmax")$ \\
$C=sqrt(A)$ & $C=sapply(A, "sqrt")$ \\
$C=abs(A)$ & $C=sapply(A, "abs")$ \\
\hline
$c=sum(A)$ & $c=agg(A, "+")$ \\
$C=rowSums(A)$ & $C=agg.row(A, "+")$ \\
$C=colSums(A)$ & $C=agg.col(A, "+")$ \\
$c=any(A)$ & $c=agg(A, "|")$ \\
$c=all(A)$ & $c=agg(A, "\&")$ \\
\hline
$C=A \%*\% B$ & $C=inner.prod(A, B, "*", "+")$ for integers; \\
 & use BLAS for floating-point values; \\
 & use SpMM \cite{SEM_SpMM} for sparse matrices. \\
\hline
\end{tabular}
\normalsize
\label{tbl:Rfuns}
\end{center}
\end{table}

In addition to computation operations, FlashR provides functions for matrix
construction and data access in a FlashR matrix (Table \ref{tbl:utility}). This
includes creating vectors and matrices, loading matrices from an external source,
interacting with the R framework, reshaping a matrix, accessing rows and columns of
a matrix and so on. FlashR avoids unnecessary data movement in these functions.
For example, transpose of a matrix does not physically move elements in the matrix
and instead, it causes FlashR to access data in the matrix differently in
the subsequent matrix operations.

\begin{table}
\begin{center}
\caption{Some of the matrix creation and matrix access functions in FlashR.}
\vspace{-10pt}
\footnotesize
\begin{tabular}{|l|l|l|}
\hline
Function & Description \\
\hline
$rep.int$ & Create a vector of a repeated value \\
$seq.int$ & Create a vector of sequence numbers \\
$runif.matrix$ & Create a uniformly random matrix  \\
$rnorm.matrix$ & Create a matrix under a normal distribution \\
\hline
$load.dense.matrix$ & Load a dense matrix to FlashR \\
$load.sparse.matrix$ & Load a sparse matrix to FlashR \\
\hline
$as.matrix$ & Convert a FlashR matrix to an R matrix \\
$fm.as.matrix$ & Convert an R matrix to a FlashR matrix \\
\hline
$t$ & Matrix transpose \\
$rbind$ & Concatenate matrices by rows \\
$cbind$ & Concatenate matrices by columns \\
$[]$ & Get rows or columns from a matrix \\
\hline
\end{tabular}
\normalsize
\label{tbl:utility}
\end{center}
\end{table}




\subsection{Dense matrices}
FlashR optimizes for dense matrices that are rectangular---with
a longer and shorter dimension---because of their frequent occurrence
in machine learning and statistics. Dense matrices can be stored
physically in memory or on SSDs or represented virtually by a sequence of
computations.

\subsubsection{Tall-and-skinny (TAS) matrices}
\begin{figure}
	\centering
	\includegraphics[scale=0.4]{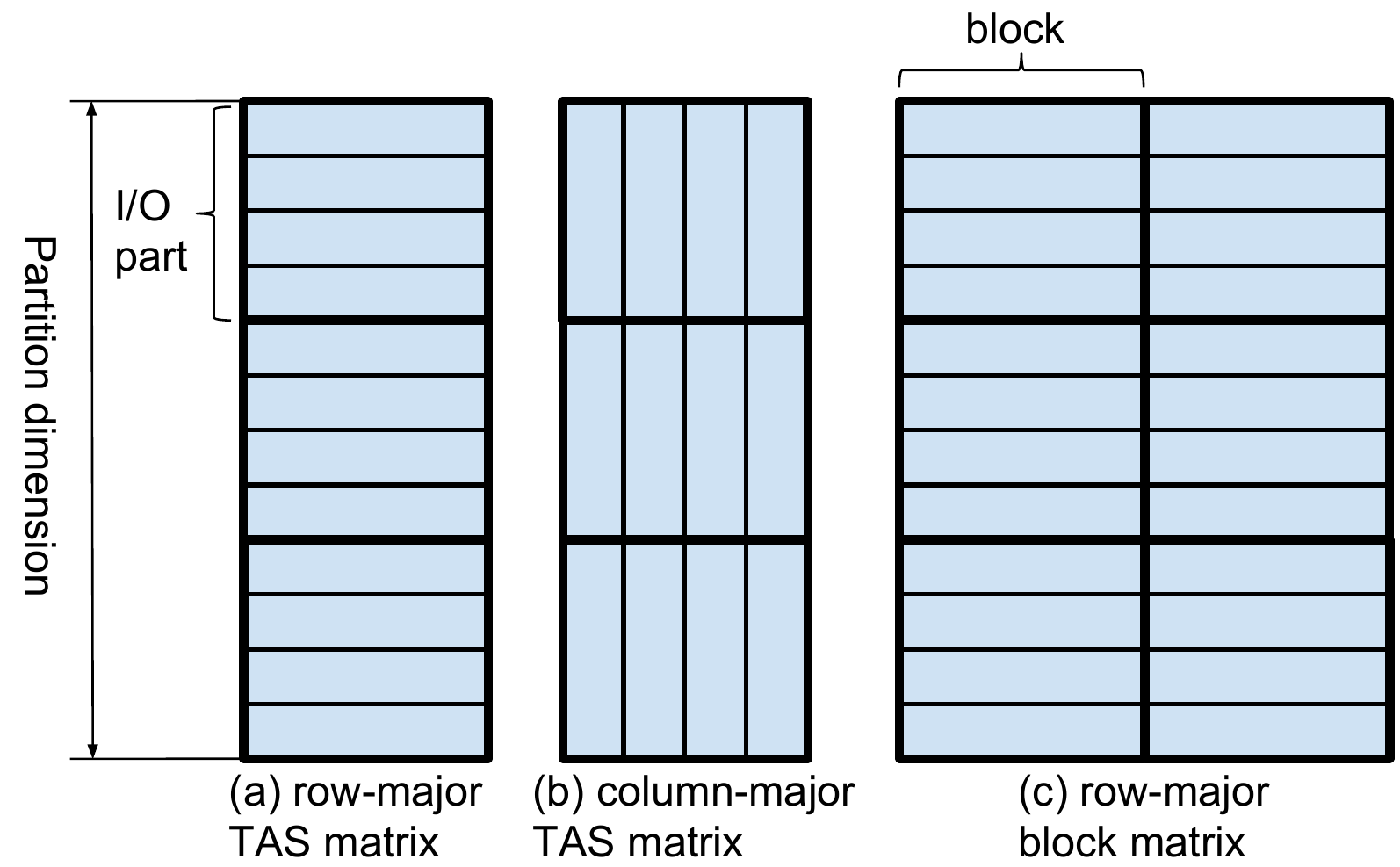}
	\vspace{-5pt}
	\caption{The format of a tall dense matrix.}
	\label{fig:den_mat}
  \vspace{-12pt}
\end{figure}

A data matrix may contain a large number of samples with a few features
(tall-and-skinny),
or a large number of features with a few samples (wide-and-short).
We use similar strategies to optimize wide-and-short matrices. FlashR
supports both row-major and column-major layouts (Figure \ref{fig:den_mat}(a)
and (b)), which allows FlashR to transpose matrices without a copy.
We store vectors as a one-column TAS matrix.

A TAS matrix is partitioned physically into I/O-partitions (Figure
\ref{fig:den_mat}). We refer to the dimension that is partitioned as
\textit{partition dimension}. All elements in an I/O-partition are stored
contiguously regardless of the data layout in the matrix. All 
I/O-partitions have the same number of rows regardless of
the number of columns in a TAS matrix. The number of rows in
an I/O-partition is always $2^i$. This produces column-major TAS
matrices whose data are well aligned in memory to help CPU vectorization.

When a TAS matrix is stored in memory, FlashR stores its I/O partitions in
fixed-size memory chunks (e.g., 64MB) across NUMA nodes.
I/O partitions from different matrices may have different sizes. By storing
I/O partitions in fixed-size memory chunks shared among all in-memory matrices,
FlashR can easily recycle memory chunks to reduce memory allocation overhead.

When a TAS matrix is stored on SSDs, it is stored as a SAFS file \cite{safs}.
For such a matrix, an I/O partition is accessed in a single I/O request.
We rely on SAFS to map the data of a matrix evenly across SSDs. SAFS allows
applications to specify the data mapping strategies. By default, FlashR uses
a hash function to map data to SSDs to fully utilize the bandwidth of all SSDs
even if we access only a subset of columns from a TAS matrix.

\subsubsection{Block matrices} \label{sec:block_mat}
FlashR stores a tall matrix as a \textit{block matrix} 
(Figure \ref{fig:den_mat}(c)) comprised of TAS blocks with $32$ columns each,
except the last block. Each block is stored as a separate TAS matrix. 
We decompose a matrix operation
on a block matrix into operations on individual TAS matrices to take advantage
of the optimizations on TAS matrices and reduce data movement.
Coupled with the I/O partitioning on TAS matrices, this strategy enables
2D-partitioning on a dense matrix and each partition fits in main memory.


\subsubsection{Virtual matrices} \label{virt_mat}
To support lazy evaluation, FlashR uses \textit{virtual matrices} that
materialize data on the fly during computation and transfer output as input to
the matrix operation. The data of a matrix are not stored physically.
All GenOps output virtual matrices. A GenOp on a \textit{block matrix} may output
a block \textit{virtual matrix}. Virtual matrices are assembled to construct
a directed acyclic graph (DAG) that represents a sequence of matrix computations
(Section \ref{dag}). Virtual matrices are essential to reduce data
movement in the memory hierarchy and memory allocation overhead to accelerate
computation.


\subsubsection{Sink matrices}
The leaf nodes in a DAG are \textit{sink matrices}.
These are produced by aggregation, groupby, and inner product GenOps for which 
the output matrices have a different \textit{partition dimension} size than
the input matrices. \textit{Sink matrices} tend to be small and we store their
results in memory. The maximum size of a sink matrix from an aggregation
is $\sqrt{N}$ for $N$ elements in the input matrix and for groupby
$k \times \sqrt{N}$ for $k$ groups, where $k$ is usually a small number.
For most of machine learning and data analysis tasks, the output of inner product
of a wide matrix with a tall matrix is small because
the long dimension of these matrices is much larger than the short dimension.


\subsection{Sparse matrices}
FlashR supports sparse matrices and optimizes sparse matrices of different
shapes differently.
For large sparse matrices that arise from graphs (e.g., social networks
and Web graphs), which have a large number of rows and columns, FlashR integrates
with our prior work \cite{SEM_SpMM} that stores sparse matrices
in a compact format on SSDs and performs sparse matrix multiplication
in semi-external memory, i.e., we keep the sparse matrix on SSDs and
the dense matrix or part of the dense matrix in memory.
Because many graph algorithms can be formulated with sparse matrix multiplication
\cite{linear_algebra}, we can express these algorithms in FlashR. In contrast,
for sparse matrices with many rows and few columns or with many columns
and few rows, FlashR stores the sparse matrices with the coordinate format
(COO). These sparse matrices can be used in sparse random projection
\cite{sparse_proj} or to store categorial values for computation.
FlashR keeps these sparse matrices in memory.




\subsection{Reducing data movement}\label{sec:datamove}
It is essential to reduce data movement within the memory hierarchy to achieve
efficiency when evaluating a sequence of matrix operations. FlashR lazily
evaluates most matrix
operations, especially the ones that do not contain heavy computation,
and constructs one or more directed-acyclic graphs (DAGs) connecting GenOps and
matrices. Figure \ref{fig:dag} (a) shows a DAG for an iteration of k-means.
It performs all computation in a DAG together, which creates opportunities
for fusing matrix operations to reduce data movement in the memory hierarchy.
This data-driven, lazy evaluation allows out-of-core problems 
to approach in-memory speed.

\subsubsection{Directed-acyclic graphs (DAGs)} \label{dag}
A DAG comprises a set of matrix nodes (rectangles) and computation nodes
(ellipses) (Figure \ref{fig:dag} (b)). The majority of matrix nodes are
virtual matrices (dashed line rectangles).
For k-means, only the input matrix \textit{X} has materialized data.
A computation node references a GenOp and input matrices and
may contain some immutable computation state, such as scalar variables and
small matrices. 

FlashR grows each DAG as large as possible, by allowing virtual matrices
of different shapes in a DAG, to increase the ratio of computation to I/O
(Figure \ref{fig:dag}). All virtual matrices in internal matrix nodes have
the same \textit{partition dimension} size to simplify evaluation and data
flow. \textit{Sink matrices} are edge nodes
in the DAG because they have different \textit{partition dimension} sizes
from the internal matrices. Any computation that uses these
sink matrices cannot be connected to the same DAG.  


\subsubsection{DAG materialization}
FlashR allows both explicit and implicit DAG materialization to simplify
programming while providing the opportunities to tune the code for
better speed. For explicit materialization, a user can invoke
\textit{fm.materialize} to materialize a virtual matrix. Access to elements
of a sink matrix triggers materialization implicitly. Materialization on
a virtual matrix triggers materialization on the DAG where the virtual
matrix connects.

By default, FlashR saves the computation results of all sink matrices of
the DAG in memory and discard the data of non-sink matrices on the fly.
Because sink matrices tend to be small, this materialization
rule leads to small memory consumption.
In exceptional cases, especially for iterative algorithms,
it is helpful to save some non-sink matrices to avoid
redundant computation and I/O across iterations.  We allow users to
set a flag on any virtual matrix to materialize and cache data in memory
or on SSDs during computation, similar to caching
a \textit{resilient distributed dataset} (RDD) in Spark \cite{spark}.

\begin{figure}
	\centering
	\includegraphics[scale=0.6]{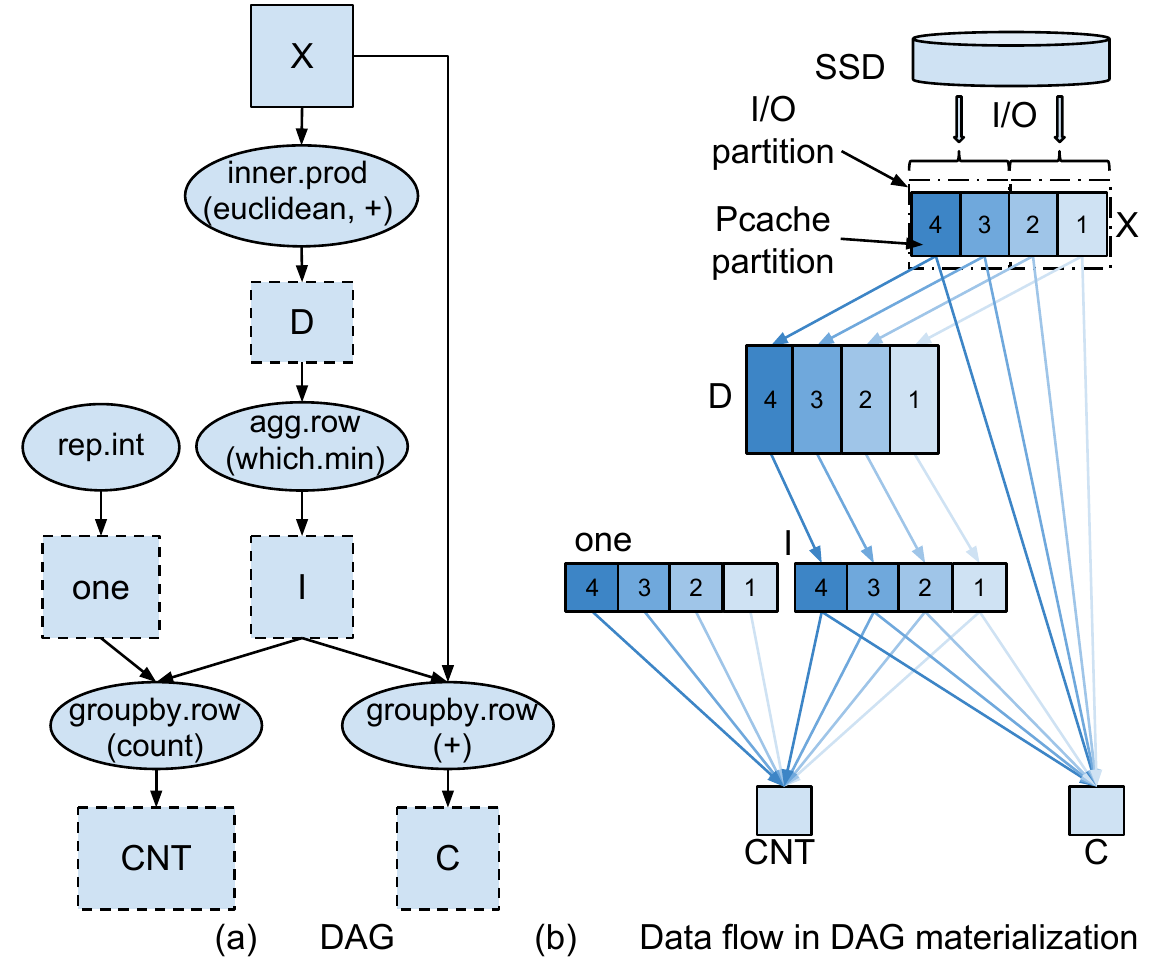}
  \vspace{-4pt}
	\caption{(a) Matrix operations are lazily evaluated to form
	a directed-acyclic graph (DAG); (b) The data flow in DAG materialization
	with two levels of partitioning: matrix X on SSDs is first partitioned
	and is read to memory in I/O partitions; an I/O partition is further
	split into processor cache (Pcache) partitions; once a Pcache partition
	is materialized, it is passed to the next GenOp to reduce CPU cache misses. }
	\label{fig:dag}
  \vspace{-8pt}
\end{figure}

FlashR partitions matrices in a DAG and materializes partitions separately.
This is possible because all matrices, except
sink matrices, share the same partition dimension. 
A partition $i$ of a virtual matrix requires data only from partitions
$i$ of the parent matrices.  All DAG operations in a partition are processed by 
the same thread so that all data required by the computations are stored and
accessed in the memory close to the processor to increase the memory bandwidth
in a NUMA machine.
When materializing a sink matrix, each thread computes partial
aggregation results on the partitions assigned to the thread. 
Then, FlashR merges per-thread partial results to construct the output.

FlashR uses two-level partitioning on dense matrices to reduce data movement
between SSDs and CPU (Figure \ref{fig:dag} (b)). It reads data on SSDs in
I/O partitions and assigns these partitions to a thread as a parallel task.
It further splits I/O-partitions into processor cache (Pcache)
partitions at run time.  Each thread materializes one Pcache-partition
at a time. Matrix operations run on TAS matrices or block matrices are
divided into TAS matrices so that a Pcache-partition fits in the CPU L1/L2 cache.

FlashR makes an effort to increase hits in the CPU cache, which
significantly reduces data movement between CPU and memory.
When materializing output to a virtual matrix, the thread passes
the output as input to the subsequent
GenOp, instead of materializing the next Pcache-partition.
This ensures that a Pcache-partition resides in the CPU cache
when the next GenOp consumes it. 
In each thread, all intermediate matrices have only one 
Pcache-partition materialized
at any time to reduce CPU cache pollution.




\subsection{Parallel execution and I/O access}
FlashR dispatches computation to threads so that they
issue large reads and writes to SSDs, while still achieving good load balancing.
FlashR uses a global task scheduler to assign I/O-partitions to threads
dynamically. Initially, the scheduler assigns multiple contiguous I/O-partitions
to a thread. The thread reads these in a single large I/O.
The number of contiguous I/O-partitions assigned to a thread is determined by
the block size of SAFS.
As the computation nears an end, the scheduler dispatches single I/O-partitions. 
The scheduler dispatches I/O-partitions sequentially to maximize contiguity in memory
and on SSD. When the DAG stores the materialized result of a non-sink matrix,
contiguity makes it easier for the file system to merge
writes from multiple threads, which helps to sustain write throughput and reduces
write amplification \cite{ripq}.




\section{Machine learning algorithms} \label{sec:apps}
To illustrate the programming interface of FlashR, we showcase some classic
algorithms written in FlashR.
Some of the algorithms can be implemented with R \textit{base} functions and
can run in the existing R framework without any modification.

\begin{figure}
\centering
\begin{lstlisting}[
	basicstyle=\footnotesize\ttfamily,
	language=R,
	tabsize=2,
	]
# X is the data matrix. C is cluster centers.
kmeans.iter <- function(X,C) {
	# Compute pair-wise distance.
	D<-inner.prod(X,t(C), "euclidean","+")
	# Find the closest center.
	I<-agg.row(D,"which.min")
	# Count the number of data points in each cluster.
	CNT<-groupby(rep.int(1,nrow(I)),I,"+")
	# Compute the new centers.
	C<-mapply.row(groupby.row(X,I,"+"),CNT,"/")
	list(C=C,I=I)
}
\end{lstlisting}
\vspace{-10pt}
	\caption{The FlashR implementation for an iteration of k-means.}
	\label{fig:kmeans}
\vspace{-10pt}
\end{figure}

K-means is a commonly used clustering algorithm that partitions data points
into $k$ clusters so that each cluster has the minimal mean distance between
the data points and the cluster center. We use this algorithm
to illustrate programming with GenOps (Figure \ref{fig:kmeans}).
It first uses \textit{inner.prod} to
compute the Euclidean distance between every data point and every cluster center
and outputs a matrix with each row representing the distances to centers.  
It uses \textit{agg.row} to find the closest
cluster for each data point.  The output matrix 
assigns data points to clusters. It then uses \textit{groupby.row} to count
the number of data points in each cluster and compute the mean of each cluster.

\begin{figure}
\begin{lstlisting}[
	basicstyle=\footnotesize\ttfamily,
	language=R,
	tabsize=2,
	]
# `X' is the data matrix, whose rows represent
# data points and columns represent features.
# `y' stores the labels of data points. `w' is
# the weight vector.
logistic.regression <- function(X,y) {
	Xtw <- X%*%t(w)
	grad <- function(X,y,w)
		(t(X)%*%(1/(1+exp(-Xtw))-y))/length(y)
	cost <- function(X,y,w)
		sum(y*(-Xtw)+log(1+exp(Xtw)))/length(y)
	# Gradient descent with line search.
	theta <- matrix(rep(0, num.features), nrow=1)
	for (i in 1:max.iters) {
		g <- grad(X, y, theta)
		l <- cost(X, y, theta)
		eta <- 1
		delta <- 0.5 * (-g) %*% t(g)
		while (cost(X, y, theta+eta*(-g)) < l+delta*eta)
			eta <- eta * 0.2
		theta <- theta + (-g) * eta
	}
	theta
}
\end{lstlisting}
\vspace{-10pt}
\caption{A simplified implementation of logistic regression using
steepest gradient descent in FlashR.}
\label{logistic}
\vspace{-5pt}
\end{figure}

Logistic regression is a commonly used classification algorithm.
We implement this algorithm solely with the R \textit{base} functions
overriden by FlashR. Figure
\ref{logistic} implements logistic regression for problems with binary-class
labels. It uses steepest gradient descent with line search to minimize
the \textit{cost} function. This example does not have a regularization term
in the cost function.

PageRank is a well-known algorithm for graph analysis. The matrix formulation
(Figure \ref{pagerank}) operates on the adjacency matrix of a graph. It uses
a power method and performs a sequence of sparse matrix multiplications until
the algorithm converges.
\begin{figure}
\begin{lstlisting}[
	basicstyle=\footnotesize\ttfamily,
	language=R,
	tabsize=2,
	]
# `graph' is a sparse matrix that represents
# the input graph. `d' is a damping factor,
# `epsilon' is the convergence accuracy.
pagerank<-function(graph,d=0.15,epsilon=1e-2){
	N <- nrow(graph)
	pr1 <- fm.rep.int(1/N, N)
	out.deg <- graph %*% fm.rep.int(1, N)
	converge <- 0
	graph <- t(graph)
	while (converge < N) {
		pr2 <- (1-d)/N+d*(graph %*% (pr1/out.deg))
		diff <- abs(pr1-pr2)
		converge <- sum(diff < epsilon)
		pr1 <- pr2
	}
	pr1
}
\end{lstlisting}
\vspace{-10pt}
\caption{A simplified implementation of PageRank in FlashR.}
\label{pagerank}
\vspace{-10pt}
\end{figure}

\section{Experimental evaluation}
We evaluate the efficiency of FlashR on statistics and machine learning
algorithms both in memory and on SSDs. We compare the R implementations of
these algorithms with the ones in two optimized parallel machine learning
libraries H2O \cite{h2o} and Spark MLlib \cite{mllib}. We further use FlashR
to accelerate existing R functions in the MASS package and compare with
Revolution R Open \cite{rro}.

We conduct experiments on a NUMA machine with four Intel Xeon E7-4860 2.6 GHz
processors and 1TB of DDR3-1600 memory. Each processor has 12 cores. The machine
is equipped with 24 OCZ Intrepid 3000 SSDs, which together are capable of
12 GB/s for read and 10 GB/s for write. The machine runs Ubuntu 14.04 and
uses ATLAS 3.10.1 as the default BLAS library.

\subsection{Benchmark algorithms}\label{benchalg}
We benchmark FlashR with some commonly used algorithms. Like the algorithms
shown in Section \ref{sec:apps}, we implement these algorithms completely with
the R code and rely on FlashR to execute them in parallel and out of core.

\noindent \textbf{Correlation} computes pair-wise Pearson's correlation
\cite{cor} and is commonly used in statistics.

\noindent \textbf{Principal Component Analysis (PCA)} computes uncorrelated
variables from a large dataset. PCA is commonly used for dimension reduction
in many data analysis tasks. We compute PCA by computing eigenvalues on the Gramian
matrix $A^T A$ of the input matrix $A$.

\noindent \textbf{Naive Bayes} is a classifier that applies Bayes' theorem
with the ``naive'' assumption of independence between every pair of features.
Our implementation assumes data follows the normal distribution.

\noindent \textbf{Logistic regression} is a linear regression model with
categorical dependent variables. We use the Newton-CG algorithm
to optimize logistic regression. It converges when
$logloss_{i-1}-logloss_i < 1e-6$, where $logloss_i$ is the logarithmic loss
at iteration $i$.

\noindent \textbf{K-means} is an iterative clustering algorithm that
partitions data points to $k$ clusters. Its R implementation is illustrated
in Figure \ref{fig:kmeans}. In the experiments, we run k-means to split
a dataset into 10 clusters by default. It converges when no data points
move.

\noindent \textbf{Multivariate Normal Distribution (mvrnorm)} generates
samples from the specified multivariate normal distribution. We use
the implementation in the MASS package.

\noindent \textbf{Linear discriminant analysis (LDA)} is a linear classifier
that assumes the normal distribution with a different mean for each class
but sharing the same covariance matrix among classes. Our LDA is adapted from
the one in the R MASS package only with some trivial modifications.

\noindent \textbf{PageRank} is a well-known algorithm for graph analysis.
We run the R implementation in Figure \ref{pagerank} in the experiment.
PageRank converges when
$\forall j, abs(PR^i(v_j) - PR^{i-1}(v_{j})) < \dfrac{0.01}{n}$,
where $PR^i(v_j)$ is the PageRank value of vertex $v_j$ at iteration $i$ and
$n$ is the number of vertices in the graph.

\begin{table}
\begin{center}
\caption{Computation and I/O complexity of the benchmark algorithms. $n$ is
	the number
	of data points, $p$ is the number of the features in a point, and $k$ is
	the number of clusters, $nnz$ is the number of non-zero entries in
	a sparse matrix. We assume $n > p$. We show the computation and I/O
	complexity of an iteration in iterative algorithms. PageRank typically
	runs on a $n \times n$ sparse matrix. Logistic regression is optimized
	with Newton-CG.
}
\vspace{-10pt}
\footnotesize
\begin{tabular}{|c|c|c|c|c|}
\hline
Algorithm & Computation & I/O \\
\hline
Correlation & $O(n \times p^2)$ & $O(n \times p)$ \\
\hline
PCA & $O(n \times p^2)$ & $O(n \times p)$ \\
\hline
Naive Bayes & $O(n \times p)$ & $O(n \times p)$ \\
\hline
Logistic regression & $O(n \times p^2)$ & $O(n \times p)$ \\
\hline
K-means & $O(n \times p \times k)$ & $O(n \times p)$ \\
\hline
mvrnorm & $O(n \times p^2)$ & $O(n \times p)$ \\
\hline
LDA & $O(n \times p^2)$ & $O(n \times p)$ \\
\hline
PageRank & $O(nnz + n)$ & $O(nnz)$ \\
\hline
\end{tabular}
\normalsize
\label{tbl:algs}
\end{center}
\vspace{-10pt}
\end{table}



These algorithms have various ratios of computation complexity and I/O complexity
(Table \ref{tbl:algs}) to thoroughly evaluate performance of FlashR on SSDs.
Logistic regression, K-means and PageRank run iteratively and, thus, we show
their computation and I/O complexity in a single iteration.

We benchmark FlashR with two real-world datasets with billions of data points
(Table \ref{tbl:data}). The Criteo dataset has over four billion data points
with binary labels (click vs. no-click), used for advertisement click
prediction \cite{criteo}. PageGraph is the adjacency matrix of a graph whose
vertices represent Web pages
and edges represents hyperlinks between Web pages \cite{webgraph}. PageGraph-32ev
are 32 singular vectors that we computed on the largest connected component of
Pagegraph with the tools we built previously \cite{flashgraph, SEM_SpMM}.
To compare FlashR with other frameworks, we take part of the Criteo and
PageGraph-32ev datasets to
create smaller datasets. PageGraph-32ev-sub is the first 336 million data points
of the PageGraph-32ev dataset. Criteo-sub contains the data points collected
on the first two days, which is about one tenth of the whole dataset.

\begin{table}
\begin{center}
\caption{Datasets. PageGraph is a sparse matrix and the other datasets are
	dense matrices.}
\vspace{-10pt}
\footnotesize
\begin{tabular}{|c|c|c|c|c|}
\hline
Data Matrix & \#rows & \#cols & \#nnz \\
\hline
PageGraph \cite{webgraph} & 3.5B & 3.5B & 128B \\
\hline
PageGraph-32ev \cite{webgraph} & 3.5B & 32 & 107B \\
\hline
Criteo \cite{criteo} & 4.3B & 40 & 172B \\
\hline
PageGraph-32ev-sub \cite{webgraph} & 336M & 32 & 11B \\
\hline
Criteo-sub \cite{criteo} & 325M & 40 & 13B \\
\hline
\end{tabular}
\normalsize
\label{tbl:data}
\end{center}
\vspace{-10pt}
\end{table}

\subsection{Comparative performance}
We evaluate FlashR against H2O \cite{h2o} and Spark MLlib \cite{mllib} as well
as Revolution R Open \cite{rro} in a large parallel machine with 48 CPU cores
and in the Amazon cloud. When running in the 48 CPU core machine, all frameworks
use 48 threads and H2O and MLlib have a large heap size (500GB) to ensure that
all data are cached in memory. When running in the cloud, we run FlashR
in a single i2.8xlarge instance (16 CPU cores, 244GB RAM and 6.4TB SSD storage)
and MLlib in a cluster of four EC2 c4.8xlarge instances (72 CPU cores, 240GB RAM
and 10Gbps network). We also use FlashR to parallelize functions (mvrnorm and LDA)
in the R MASS package and compare their speed with Revolution R Open. We use
Spark v2.0.1, H2O v3.10.2 and Revolution R Open v3.3.2.

\begin{figure}
  \vspace{-5pt}
	\centering
	\footnotesize
	\begin{subfigure}{.5\textwidth}
		\includegraphics{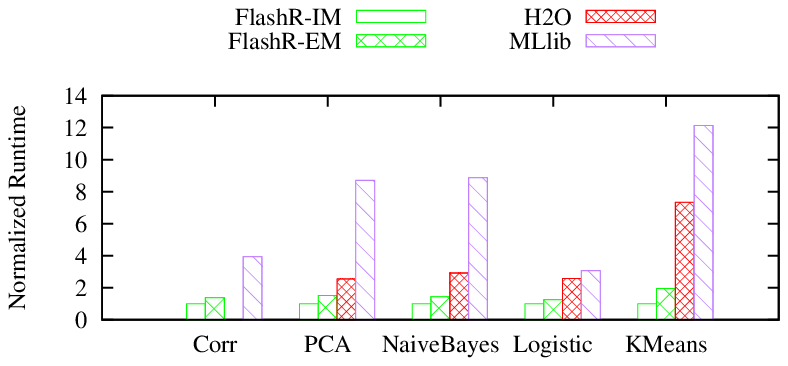}
		\caption{In a large parallel machine with 48 CPU cores.}
		\label{perf:para}
	\end{subfigure}

	\vspace{3pt}
	\begin{subfigure}{.5\textwidth}
		\includegraphics{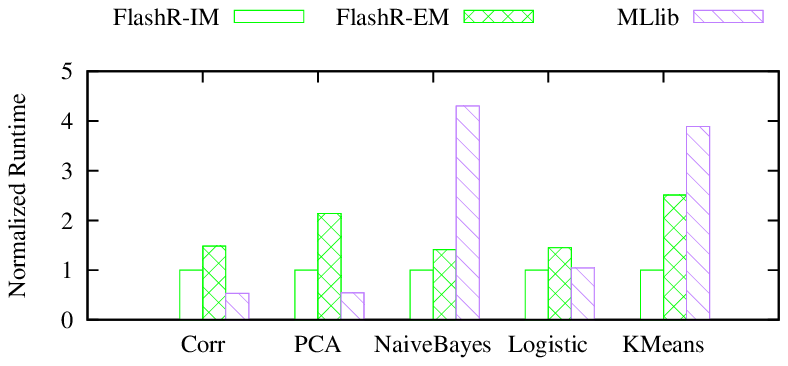}
		\caption{In the Amazon cloud. FlashR-IM and FlashR-EM run on one
			EC2 i2.8xlarge instance (16 CPU cores) and Spark MLlib runs
		on a cluster of four EC2 c4.8xlarge instances (72 CPU cores).}
		\label{perf:cloud}
	\end{subfigure}
	\vspace{-8pt}
	\caption{The normalized runtime of FlashR in memory (FlashR-IM) and
	on SSDs (FlashR-EM) compared with H2O and Spark MLlib. Correlation is not
	available in H2O. We run k-means on the PageGraph-32ev-sub dataset and
	all other algorithms on the Criteo-sub dataset.}
	\label{perf:rt}
  \vspace{-10pt}
\end{figure}

FlashR outperforms H2O and Spark MLlib significantly on all algorithms
(Figure \ref{perf:para}) in the large parallel machine with 48 CPU cores.
FlashR running in memory achieves 3 to 10 times performance gain when compared
with MLlib, and 2.5 to 7 times performance gain when compared with H2O.
When running on SSDs, FlashR achieves at least half the speed of running in
memory. Even though logistic regression in FlashR uses Newton-CG, a much more
computation intensive optimization algorithm than LBFGS \cite{lbfgs} used by
H2O and MLlib (Newton-CG takes 10 iterations and LBFGS takes 14 iterations
to converge to similar loss on the dataset), FlashR still runs almost $2-3$
times as fast as H2O and MLlib. All implementations rely on BLAS for
matrix multiplication. H2O and MLlib have to
implement non-BLAS operations with Java and Scala, respectively, and
MLlib materializes operations such as aggregation separately. In contrast,
FlashR fuses matrix operations and performs two-level partitioning to
minimize data movement in the memory hierarchy and keeps data in local
memory to achieve high memory bandwidth.

We further evaluate the speed of FlashR on Amazon EC2 cloud and compare it with
Spark MLlib on an EC2 cluster (Figure \ref{perf:cloud}). Spark MLlib needs
at least 4 c4.8xlarge instances to process the datasets (PageGraph-32ev-sub
and Criteo-sub).
Even though Spark MLlib has 4.5 as much computation power as FlashR, FlashR
at least matches the speed of Spark MLlib and even outperforms it.

\begin{figure}[b]
  \vspace{-10pt}
	\begin{center}
		\footnotesize
		\includegraphics{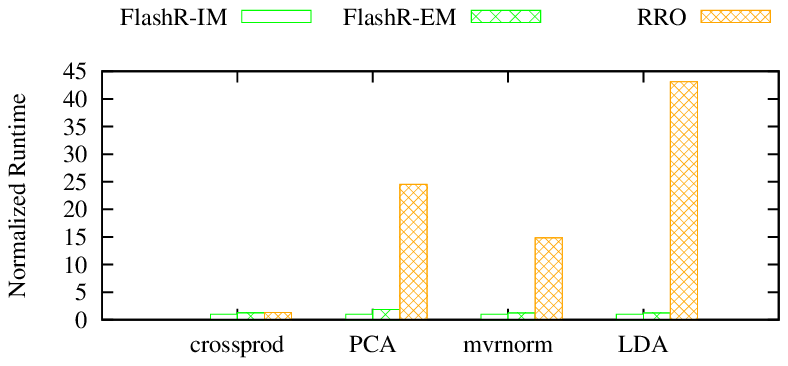}
		\vspace{-10pt}
		\caption{In-memory (FlashR-IM) and out-of-fore (FlashR-EM) FlashR
		compared with Revolution R Open on a data matrix with one million rows
		and one thousand columns when running on the parallel machine with
		48 CPU cores.}
		\label{fig:fmR}
	\end{center}
  \vspace{-15pt}
\end{figure}

FlashR running both in memory and on SSDs outperforms Revolution R Open by more
than an order of magnitude even on a small dataset ($n=1,000,000$ and $p=1000$)
(Figure \ref{fig:fmR}).
Revolution R Open uses Intel MKL to parallelize matrix multiplication. As such,
we only compare the two frameworks with computations that use matrix
multiplication heavily. Both FlashR and Revolution R Open run the mvrnorm
and LDA implementations from the MASS package. For simple matrix operations such as crossprod,
FlashR slightly outperforms Revolution R Open. For more complex computations,
the speed gap between FlashR and Revolution R increases. Even though matrix multiplication
is the most computation-intensive operation in an algorithm, it is insufficient
to only parallelize matrix multiplication to achieve high efficiency.

\subsection{Scalability}

We show the scalability of FlashR on the billion-scale datasets in Table
\ref{tbl:data}. In these experiments, we run the iterative algorithms on
the datasets until they converge (see their convergence condition in Section
\ref{benchalg}).

\begin{table}
\begin{center}
	\caption{The runtime and memory consumption of FlashR on the billion-scale
		datasets on the 48 CPU core machine. The runtime of iterative
		algorithms is measured when the algorithms converge. We run PageRank
		on the PageGraph dataset, run k-means on PageGraph-32ev and the remaining
		algorithms on Criteo.}
\vspace{-10pt}
\footnotesize
\begin{tabular}{|c|c|c|}
\hline
	& Runtime (s) & Memory (GB) \\
\hline
Correlation & $91.23$ & $1.5$ \\
\hline
PCA & $136.71$ & $1.5$ \\
\hline
NaiveBayes & $76.55$ & $3$ \\
\hline
LDA & $2280$ & $8$ \\
\hline
Logistic regression & $4154.40$ & $26$ \\
\hline
k-means & $1110.82$ & $28$ \\
\hline
PageRank & $3900$ & $135$ \\
\hline
\end{tabular}
\normalsize
\label{tbl:scale}
\end{center}
\vspace{-10pt}
\end{table}

Even though we process the billion-scale datasets in a single machine, none of
the algorithms are prohibitively expensive. Simple algorithms, such as
Naive Bayes and PCA, require one or two passes over the datasets and take
only one or two minutes to complete. Logistic regression and k-means take
about $10-20$ iterations to converge. Because the PageRank implementation
uses the power method, it takes $100$ iterations to converge.
Nevertheless, all of the iterative algorithms take about one hour or less.

FlashR scales to datasets with billions of data points easily when running
out of core. Most of the algorithms have negligible memory consumption.
PageRank consumes more memory because the sparse matrix multiplication in
PageRank keeps vectors in memory for semi-external memory computation.
The scalability of FlashR is mainly bound by the capacity of SSDs.
The functional programming
interface generates a new matrix in each matrix operation, which potentially
leads to high memory consumption. Thanks to lazy evaluation and virtual matrices,
FlashR only needs to materialize the small matrices to effectively reduce
memory consumption.

\subsection{Computation complexity versus I/O complexity}
We further compare the speed of FlashR in memory and in external memory
for algorithms with different computation and I/O complexities.
We pick three algorithms from Table \ref{tbl:algs}: \textit{(i)} Naive Bayes,
whose computation and I/O complexity are the same, \textit{(ii)}
correlation, whose computation complexity grows quadratically with $p$ while
its I/O complexity grows linearly with $p$, \textit{(iii)} k-means, whose computation
complexity grows linearly with $k$ while its I/O complexity is independent
to $k$. We run the first two algorithms on datasets with $n=100M$ and $p$
varying from 8 to 512. We run k-means on a dataset with $p=100M$ and $p=32$
and vary the number of clusters from 2 to 64.

\begin{figure}[t]
	\begin{center}
		\footnotesize
		\includegraphics{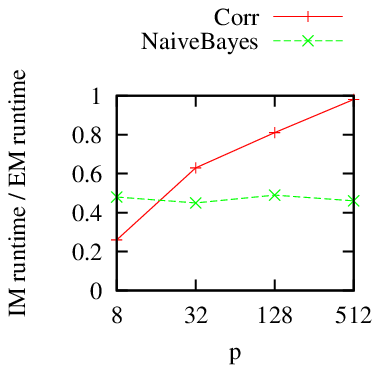}
		\includegraphics{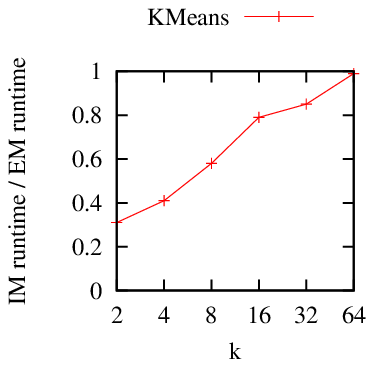}
		\vspace{-10pt}
		\caption{The relative runtime of FlashR in memory versus on SSDs
		on a dataset with $n=100M$ while varying $p$ (the number of features)
		on the left and varying $k$ (the number of clusters) on the right.}
		\label{perf:stat}
	\end{center}
  \vspace{-15pt}
\end{figure}

As the number of features or clusters increases, the performance gap between
in-memory and external-memory execution narrows and the external-memory
performance approaches in-memory performance for correlation and k-means
but not Naive Bayes (Figure \ref{perf:stat}). This observation conforms with
the computation and I/O complexity of the algorithms in Table \ref{tbl:algs}.
For correlation and k-means, the number of clusters or features causes computation
to grow more quickly than the I/O, driving performance toward a computation bound.
The computation bound can be realized on few features or clusters for an I/O throughput of 10GB/s.
Because most of the machine learning algorithms in Table \ref{tbl:algs} have
computation complexities that grow quadratically with $p$, we expect FlashR on SSDs to
achieve the same speed as in memory on datasets with a higher dimension size.

\section{Conclusions}
We present FlashR, a matrix-oriented programming framework that executes
R-programmed machine learning algorithms in parallel and out-of-core
automatically. FlashR scales to large datasets by utilizing commodity SSDs.


Although R is considered to be slow and unable to scale to large datasets,
we demonstrate that with sufficient system-level optimizations, FlashR powers
the R programming interface to achieve high performance and scalability
for developing many machine learning algorithms. R implementations executed in FlashR
outperform H2O and Spark MLlib on all algorithms by a large factor, using
the same shared memory hardware. FlashR scales to datasets with billions of
data points easily with negligible amounts of memory and completes all
algorithms within a reasonable amount of time.

Even though SSDs are an order of magnitude slower than DRAM, the external-memory
execution of many algorithms in FlashR achieve performance approaching their in-memory
execution. We demonstrate that an I/O throughput of 10 GB/s saturates the CPU for many
algorithms, even in a large parallel NUMA machine. 

FlashR simplifies the programming effort of writing parallel and out-of-core
implementations for large-scale machine learning. With FlashR, machine learning
researchers can prototype algorithms in a familiar programming environment,
while still getting efficient and scalable implementations.
We believe FlashR provides new opportunities for developing large-scale
machine learning algorithms.


{\footnotesize \bibliographystyle{acm}
\bibliography{kdd17}}

\end{document}